\documentclass[aps,pra,onecolumn,superscriptaddress]{revtex4-2}
\pdfoutput=1
\usepackage{amssymb}
\usepackage[dvips]{graphicx}
\usepackage{enumerate}
\usepackage{epsfig}
\usepackage{diagbox}
\usepackage{xcolor}
\usepackage[T1]{fontenc}
\usepackage{amsthm,amsfonts,amssymb,amscd,mathrsfs,eufrak,xspace} 
\usepackage{amsmath}
\usepackage{color}
\usepackage{setspace}
\usepackage{url}
\usepackage{enumitem}
\usepackage[colorlinks,linkcolor=red,anchorcolor=blue,citecolor=red]{hyperref}
\usepackage{braket} 

\begin{document}


\title{Twin-field quantum key distribution with partial phase postselection}



\author{Yao Zhou}
\author{Zhen-Qiang Yin}
\email{yinzq@ustc.edu.cn}
\author{Rui-Qiang Wang}
\author{Shuang Wang}
\email{wshuang@ustc.edu.cn}
\author{Wei Chen}
\author{Guang-Can Guo}
\author{Zheng-Fu Han}
\affiliation{CAS Key Laboratory of Quantum Information, University of Science and Technology of China, Hefei, Anhui 230026, China}
\affiliation{CAS Center for Excellence in Quantum Information and Quantum Physics, University of Science and Technology of China, Hefei, Anhui 230026, China}
\affiliation{Hefei National Laboratory, University of Science and Technology of China, Hefei 230088, China}
\affiliation{State Key Laboratory of Cryptology, P. O. Box 5159, Beijing 100878, China}




\begin{abstract}
Quantum key distribution (QKD) allows two remote parties to share information-theoretically secure keys. In recent years, a revolutionary breakthrough called twin-field (TF) QKD has been developed to overcome the linear key-rate constraint and greatly increases the achievable distance. Phase-randomization and subsequent postselection play important roles in its security proof. Later, no-phase-postselection TF-QKD was proposed and became a popular variant, since the removal of phase postselection leads to a higher key rate. However, the achievable distance is decreased compared to the original one. Here, we propose a TF-QKD protocol with partial phase postselection.  Namely, its code mode is still free from global phase randomization and postselection to make sure the advantage of the high key rate remain. On other hand, phase postselection is introduced in the decoy mode to improve the performance.  Applying an operator dominance condition, we prove universal security of the proposed protocol in the finite-key case under coherent attacks,  and numerical simulations confirm its potential advantages in terms of key rate and achievable distance.
    
\end{abstract}


\maketitle

\section{introduction}
Quantum key distribution (QKD) \cite{BB84,PhysRevLett.67.661}, a practical and interesting application of quantum information science at this stage, provides information-theoretically secure keys between two distant parties, called Alice and Bob. The security of QKD does not depend on computational complexity, but is guaranteed by the fundamental laws of physics, so any eavesdropper Eve cannot guess the key with probability more than $\epsilon$ that can be chosen arbitrarily close to 0 even if she has unlimited computing resources, full access to the quantum channel between Alice and Bob, and can also read the classical messages exchanged between Alice and Bob. Over the past three decades, QKD has attracted tremendous attention and developed rapidly both in theory \cite{RN155, doi:10.1126/science.283.5410.2050, PhysRevLett.85.441,PhysRevLett.95.080501,PhysRevLett.91.057901,PhysRevLett.94.230503,PhysRevLett.94.230504,PhysRevLett.108.130503,PhysRevLett.108.130502,RN110,RN167,RN74} and experiment \cite{PhysRevLett.98.010503,PhysRevLett.98.010504,PhysRevLett.98.010505,Wang:12,PhysRevLett.111.130502,PhysRevLett.113.190501,RN113,RN131,PhysRevLett.115.160502,PhysRevLett.117.190501,PhysRevX.6.011024,RN159,Wang:17,RN114,Liu:18,RN132,RN133,PhysRevLett.123.100505,PhysRevLett.123.100506,RN134,PhysRevLett.124.070501,RN138,RN139,RN140,PhysRevLett.126.250502,RN141,RN97}. The security of QKD with ideal devices, e.g. single photon sources, has been rigorously proven from the perspective of entanglement purification \cite{RN155, doi:10.1126/science.283.5410.2050, PhysRevLett.85.441} or information theory \cite{PhysRevLett.95.080501}. But security in an ideal case does not guarantee security in practice. We must consider the practical security due to the imperfection of real devices \cite{scarani2014black,RN48}. The decoy-state method \cite{PhysRevLett.91.057901,PhysRevLett.94.230503,PhysRevLett.94.230504} guarantees security with a practical phase-randomized weak coherent optical source. MDI-QKD \cite{PhysRevLett.108.130503} (see also \cite{PhysRevLett.108.130502}) can solve all possible loopholes of detection and guarantee security even if the measurement device is controlled by Eve. During the same period, various types of QKD protocols have also been implemented \cite{PhysRevLett.98.010503,PhysRevLett.98.010504,PhysRevLett.98.010505,Wang:12,PhysRevLett.111.130502,PhysRevLett.113.190501,RN113,RN131,PhysRevLett.115.160502,PhysRevLett.117.190501,PhysRevX.6.011024,RN159,Wang:17,RN114,Liu:18}. However, all the protocols or experiments mentioned above can't break the PLOB bound \cite{RN88} (see also the TGW bound \cite{RN115}), which limits the achievable distance of QKD. 
	
	In 2018, M. Lucamarini et al. proposed the original twin-field (TF) QKD \cite{RN74} to overcome the fundamental rate-distance limit of QKD without needing quantum repeaters. TF-QKD is a special variant of MDI-QKD which also introduces an untrusted measurement station located in the channel. The two communication parties prepare pairs of weak coherent states independently and interference detection is performed at an untrusted measurement site. Specifically, the coherent pulse interference and single-photon detection are the key reasons which make the key rate scale as the square root of the channel loss rather than a linear dependence. Once it was proposed, it has received extensive attention due to the advantages of measurement-device-independence and significant improvement on achievable distances. Soon after, the security of TF-QKD was strictly proved in \cite{PhysRevX.8.031043}. Since then, various variants \cite{PhysRevA.98.062323,RN128,RN90,RN129,RN77,RN76,RN78,RN166,RN81} of TF-QKD have emerged and impressive experimental advances \cite{RN132,RN133,PhysRevLett.123.100505,PhysRevLett.123.100506,RN134,PhysRevLett.124.070501,RN138,RN139,RN140,PhysRevLett.126.250502,RN141,RN97} have also been achieved. Roughly speaking, the original TF-QKD \cite{RN74,PhysRevX.8.031043} consists of code mode and decoy mode.  In the former one, the phase $0$ or $\pi$ of a weak coherent pulse is modulated to encode raw key bit $0$ or $1$ respectively. The latter one is similar but with different intensities, the function of which is to monitor security. Additionally, a continuous random phase $\theta_a$ $(\theta_b)$ from $[0,2\pi)$ is applied in each optical pulse by Alice (Bob). Alice and Bob send their optical pulses to the third party, Charlie, who may be an agent of Eve and is expected to record and announce the detector ($L$ or $R$) clicks by performing interferometric measurements. After receiving the announcement from Charlie, Alice and Bob postselect the cases satisfying $\theta_a\approx\theta_b$ to generate sifted key bits. This phase randomization and postselection play important roles in security, but obviously reduce the key rate per trial. As an alternative, the variant called no-phase-postselection (NPP) \cite{RN128,RN90,RN129} TF-QKD removes the phase randomization in code mode, thus its key rate will be free of reduction due to phase postselection. In its decoy mode,  phase-randomization remains but postselection is also bypassed for simplicity. Although NPP-TF-QKD has a higher key rate and has become popular, its achievable distance is shorter than the original TF-QKD.  In short, there are two reasons for the shortening of distance: the  removal of phase randomization in code mode allows the eavesdropper to obtain key information more easily; and  the missing of phase information and postselection in decoy mode also makes the estimation of information leakage more loose.

	Intuitively,  phase postselection inevitably reduces the key rate, but it seems helpful for Alice and Bob to monitor the security. Inspired by this idea,  we propose a variant of the TF-type protocol with partial phase postselection. In our protocol, code mode can be the same as NPP-TF-QKD, thus still free from global phase randomization and postselection, which guarantees the advantage of the high key rate remains.  Conversely, the phase postselection is introduced in the decoy mode to help Alice and Bob to bound information leakage more accurately.  Note that decoy mode is not used to generate the key, thus the key rate is not influenced by this postselection. Based on the operator dominance method \cite{RN87}, a security proof in the finite-key region of our protocol is presented. Numerical simulations show that both the key rate and achievable distance can be significantly improved compared to the NPP-TF-QKD.

	The paper is arranged as follows. In Sec. \ref{Protocol description}, we describe the protocol in detail. In Sec. \ref{Methods}, we give a security proof in the finite-key case and show how to calculate the key rate. In Sec. \ref{Results}, numerical simulations are given to show the advantages of our protocol, and we also apply our idea in the optimized four-phase twin-field protocol \cite{RN97}. Finally, a conclusion is given.

\section{Protocol description} \label{Protocol description}
\begin{figure}
		\centering
		\includegraphics[width=\textwidth]{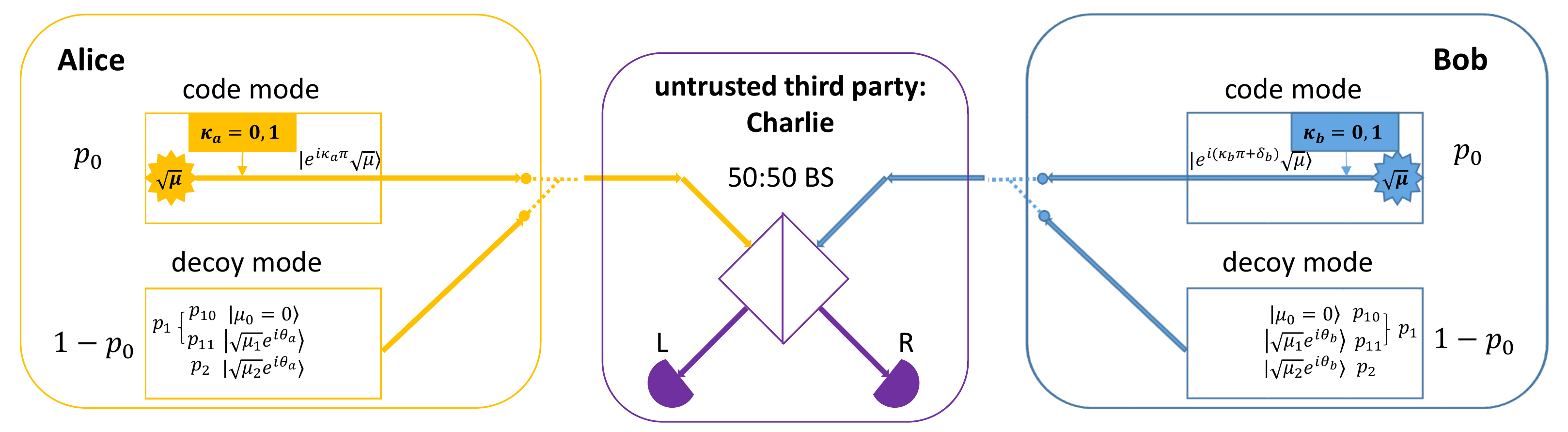}
		\caption{Illustration of our protocol. In code mode, Alice and Bob both encode their random bit in the phase of the weak coherent state with intensity $\mu$. Bob introduces the phase fluctuation $\delta \in[-\frac{\Delta}{2},\frac{\Delta}{2}]$ to meet the security proof. In decoy mode, they independently prepare nothing or weak coherent states. When they send weak coherent states, they randomize the optical phase $\theta$ and switch among two intensities $\mu_1$ and $\mu_2$. The third party, Charlie, who may be controlled by an adversary, announces the click results of the two detectors L and R. When only the right detector clicks in code mode, Bob flips his bit. After $N_{tot}$ rounds, Alice and Bob announce the intensity and phase information in decoy mode and perform corresponding post-processing to share the secret key.}
		\label{fig.protocol}
	\end{figure}

Our protocol is shown in Fig.\ref{fig.protocol}. Alice and Bob randomly choose decoy and code mode. In each mode, they send weak coherent optical pulses to Charlie, the central untrusted station. Charlie (he can be dishonest) performs interferometric measurements and announces the measurement results to Alice and Bob. The specific protocol process is as follows. 
	
	Step 0: Alice and Bob agree on the parameters $\mu$, $\mu_1$, $\mu_2$, $p_0$, $p_{10}$, $p_{11}$, $p_2$ and $\Delta$, the meanings of which are shown in the following steps. 
	All the parameters are positive and $p_0+p_{10}+p_{11}+p_2=1$. The remaining constraints on the parameters are shown in Eqs. (\ref{constraint1}), (\ref{constraint2}), and (\ref{constraint3}) in Sec. \ref{Methods}.
	
	Step 1: Alice and Bob randomly choose code mode and decoy mode with probabilities $p_0$ and $1-p_0$, respectively. In each mode, they send weak coherent pulses independently to the untrusted third party Charlie.
	
	Step 2: In the code mode, Alice (Bob) randomly generates a key bit $\kappa_a (\kappa_b) \in \{0,1\}$ and prepares the corresponding weak coherent state $\ket{e^{i\kappa_{a(b) }\pi}\sqrt{\mu}}$ with intensity $\mu$. Additionally, Bob introduces a randomized phase $\delta_b \in[-\frac{\Delta}{2},\frac{\Delta}{2}]$ which is chosen randomly according to the uniform distribution on the segment (In step 4, one can see that $\Delta/\pi$ represents the probability that the phase postselection condition is satisfied in the decoy mode, and to accumulate sufficient data in 
	 the actual experiment a typical value of $\Delta$ may be $\pi/8$.).
	In the decoy mode, Alice (Bob) prepares nothing (vacuum state $\mu_0=0$) with conditional probability in the decoy mode $p_{10}/(1-p_0)$ or a weak coherent state with intensity randomly choosing from $\mu_1$ and $\mu_2$ with probabilities $p_{11}/(1-p_0)$ and $p_2/(1-p_0)$. For Alice and Bob, they modulate a phase $\theta_{a(b)}$ on their own weak coherent state for both intensities $\mu_1$ and $\mu_2$, where $\theta_{a(b)}$ is a randomized phase chosen uniformly from $[0,2\pi)$. 
	
	Step 3: Charlie makes the two incoming states interfere on a beamsplitter (BS), two photon detectors L and R are located at its two distinct outputs. He records which detector clicks.
	
	Step 4: Steps 1 to 3 will be repeated $N_{\text{tot}}$ times. When the quantum communication is over, Charlie publicly announces all the information about the detection events. Only the trials where just one of the two detectors clicked are defined as successful trials and retained for further processing, all the other trials are discarded. Alice and Bob announce the intensities for each remaining trial. When Alice and Bob both choose the code mode (both send the intensity $\mu$), they record their key bits $\kappa_a$ and $\kappa_b$ sequentially to form the sifted key string. Note that if the click of the right detector R was announced, Bob should flip his corresponding key bit $\kappa_b$. Thus, Alice and Bob can accumulate shared random bits by repetition, which we call sifted keys. When Alice and Bob both choose the decoy mode, they retain the trials in which they both send a vacuum state. For other cases in decoy mode, they announce their random phases and only the trials that meet both of the following conditions will be retained: (1) both of them choose the same intensity, (2) phase postselection condition: the random phases $\theta_a$ and $\theta_b$ satisfy $|\theta_a-\theta_b|\ {\rm mod}\ \pi \le\frac{\Delta}{2}$. We denote the sifted key in Alice's hand as $Z_s$, the sifted key in Bob's hand as $Z_s^{\prime}$ and the length of the sifted key string as $K_0$ ($K_0$ is also the number of successful trials in the code mode). We also denote the numbers of the retained trials in which Alice and Bob both send the intensities $\mu_0$, $\mu_1$ or $\mu_2$ as $K_{10}$, $K_{11}$ and $K_2$, respectively, and $K_1\equiv K_{10}+K_{11}$.
	
	Step 5: According to $K_0$, $K_1$ and $K_2$, Alice and Bob can share a secret key string with length $G$ from their sifted key string $K_0$ after error correction and privacy amplification \cite{RennerPhD} with a failure probability no larger than $\epsilon_{\mathrm{sec}}$ (see Methods section for the proof). Error correction works as follows: Alice sends Bob $H_{\mathrm{EC}}$ bits of error correction data to correct $Z_s^{\prime}$, and Bob obtains an estimate $\hat{Z_s}$ of $Z_s$ from $Z_s^{\prime}$. Then, Alice computes a bit string (a hash) of length $\zeta^{\prime}$ by applying a random universal$_2$ hash function \cite{CARTER1979143} to $Z_s$. She sends the hash and the hash function to Bob. If the hash that Bob computes from $\hat{Z_s}$ is not the same as that of Alice, the protocol aborts. A privacy amplification scheme based on universal$_2$ hash function \cite{476316,10.1007/978-3-540-30576-7_22} finally allows Alice and Bob to extract final secret key string of length $G$.

\section{Methods} \label{Methods}
	We associate the amount of information leakage to the phase error rate $e_{\mathrm{ph}}$ in an equivalent protocol in which Alice and Bob use local auxiliary qubits A and B as a substitution for the classically controlled state preparation, which depends on the random number $\kappa_{a,b}$. Alice and Bob's procedure in code mode can be equivalently executed by preparing the qubits AB and the optical pulses ${\rm C_AC_B}$ in an entangled state 
	\begin{equation}
		\frac{\ket{0}_\mathrm{A}\ket{\sqrt{\mu}}_{\mathrm{C_A}}+\ket{1}_\mathrm{A}\ket{-\sqrt{\mu}}_{\mathrm{C_A}}}{\sqrt{2}}\otimes\frac{\ket{0}_\mathrm{B}\ket{\sqrt{\mu}e^{i\delta_b}}_{\mathrm{C_B}}+\ket{1}_\mathrm{B}\ket{-\sqrt{\mu}e^{i\delta_b}}_{\mathrm{C_B}}}{\sqrt{2}}.
	\end{equation} 
In the equivalent scenario, Alice and Bob send the optical pulses to Charlie for interferometric measurements and then measure the local auxiliary qubits in the $Z=\{\ket{0},\ket{1}\}$ or $X=\{\ket{+}=(\ket{0}+\ket{1})/\sqrt{2}\ ,\ket{-}=(\ket{0}-\ket{1})/\sqrt{2}\}$ basis. There are $K_0$ successful trials in the actual protocol where Alice and Bob both chose the code mode after sending and measuring all the coherent states. So in the equivalent protocol, if they all measure the local qubits in the Z basis, they will also obtain a sifted key of length $K_0$ among these $K_0$ successful trials. To evaluate the secrecy of the key bits, one can resort to complementarity \cite{RN127}, that is, to estimate the phase error rate $e_{\mathrm{ph}}$. Imaging that AB in these $K_0$ trials are all measured in the X basis instead of Z basis, Alice and Bob will obtain $K_0$ virtual bits. The phase error rate $e_{\mathrm{ph}}$ is defined as the bit error rate of these virtual bits. 

To evaluate $e_{\mathrm{ph}}$, let's rewrite the entangled state of the qubits AB and the optical pulses ${\rm C_AC_B}$ as
\begin{align}
	&\frac{\frac{\ket{+}_\mathrm{A}+\ket{-}_\mathrm{A}}{\sqrt{2}}\ket{\sqrt{\mu}}_{\mathrm{C_A}}+\frac{\ket{+}_\mathrm{A}-\ket{-}_\mathrm{A}}{\sqrt{2}}\ket{-\sqrt{\mu}}_{\mathrm{C_A}}}{\sqrt{2}}\otimes\frac{\frac{\ket{+}_\mathrm{B}+\ket{-}_\mathrm{B}}{\sqrt{2}}\ket{\sqrt{\mu}e^{i\delta_b}}_{\mathrm{C_B}}+\frac{\ket{+}_\mathrm{B}-\ket{-}_\mathrm{B}}{\sqrt{2}}\ket{-\sqrt{\mu}e^{i\delta_b}}_{\mathrm{C_B}}}{\sqrt{2}}\nonumber\\
	&=\bigg[\ket{+}_\mathrm{A}\frac{\ket{\sqrt{\mu}}_{\mathrm{C_A}}+\ket{-\sqrt{\mu}}_{\mathrm{C_A}}}{2}+\ket{-}_\mathrm{A}\frac{\ket{\sqrt{\mu}}_{\mathrm{C_A}}-\ket{-\sqrt{\mu}}_{\mathrm{C_A}}}{2}\bigg]\nonumber\\
	&\otimes\bigg[\ket{+}_\mathrm{B}\frac{\ket{\sqrt{\mu}e^{i\delta_b}}_{\mathrm{C_B}}+\ket{-\sqrt{\mu}e^{i\delta_b}}_{\mathrm{C_B}}}{2}+\ket{-}_\mathrm{B}\frac{\ket{\sqrt{\mu}e^{i\delta_b}}_{\mathrm{C_B}}-\ket{-\sqrt{\mu}e^{i\delta_b}}_{\mathrm{C_B}}}{2}\bigg],
\end{align}
from which, no matter Charlie's announcement, Alice always measures $\ket{+}$ with marginal probability $p_{+}=|(\ket{\sqrt{\mu}}+\ket{-\sqrt{\mu}})/2|^2=e^{-\mu}\cosh\mu$ or $\ket{-}$ with marginal probability $p_{-}=|(\ket{\sqrt{\mu}}-\ket{-\sqrt{\mu}})/2|^2=e^{-\mu}\sinh\mu$ in the X basis and the same for Bob. Note that the measurement
results of the auxiliary particles and their probability distributions in the X basis are independent of the measurement order of the auxiliary particles and the optical pulses sent.
Evidently, if Alice and Bob observe $A=B$ in the X basis measurement,
the subsystem of the original entangled state, optical pulses ${\rm C_AC_B}$, collapses into  the normalized quantum state $\rho_{\mathrm{even}}$, the probability of which is $p_{\mathrm{even}}$, i.e.
\begin{align}
	p_{\mathrm{even}}\rho_{\mathrm{even}}
	&=\frac{1}{\Delta}\int_{-\frac{\Delta}{2}}^{\frac{\Delta}{2}}\bigg[
	P\{\frac{\ket{\sqrt{\mu}}_{\mathrm{C_A}}+\ket{-\sqrt{\mu}}_{\mathrm{C_A}}}{2}\otimes\frac{\ket{\sqrt{\mu}e^{i\delta_b}}_{\mathrm{C_B}}+\ket{-\sqrt{\mu}e^{i\delta_b}}_{\mathrm{C_B}}}{2}\}\nonumber\\
	&+P\{\frac{\ket{\sqrt{\mu}}_{\mathrm{C_A}}-\ket{-\sqrt{\mu}}_{\mathrm{C_A}}}{2}\otimes\frac{\ket{\sqrt{\mu}e^{i\delta_b}}_{\mathrm{C_B}}-\ket{-\sqrt{\mu}e^{i\delta_b}}_{\mathrm{C_B}}}{2}\}\bigg]\ \mathrm{d}\delta_b
	\nonumber\\
	&=\frac{1}{\Delta}\int_{-\frac{\Delta}{2}}^{\frac{\Delta}{2}}\frac{1}{8}\ \bigg[P\{\ket{\sqrt{\mu}}_{\mathrm{C_A}}\ket{\sqrt{\mu}e^{i\delta_b}}_{\mathrm{C_B}}+\ket{-\sqrt{\mu}}_{\mathrm{C_A}}\ket{-\sqrt{\mu}e^{i\delta_b}}_{\mathrm{C_B}}\}\nonumber\\
	&+P\{\ket{\sqrt{\mu}}_{\mathrm{C_A}}\ket{-\sqrt{\mu}e^{i\delta_b}}_{\mathrm{C_B}}+\ket{-\sqrt{\mu}}_{\mathrm{C_A}}\ket{\sqrt{\mu}e^{i\delta_b}}_{\mathrm{C_B}}\}\ \bigg]\ \mathrm{d}\delta_b
\end{align}
and
\begin{align}
	p_{\mathrm{even}}&=\mathrm{tr}\bigg[\frac{1}{8}P\{\ket{\sqrt{\mu}}_{\mathrm{C_A}}\ket{\sqrt{\mu}e^{i\delta_b}}_{\mathrm{C_B}}+\ket{-\sqrt{\mu}}_{\mathrm{C_A}}\ket{-\sqrt{\mu}e^{i\delta_b}}_{\mathrm{C_B}}\}\nonumber\\
	&+\frac{1}{8}P\{\ket{\sqrt{\mu}}_{\mathrm{C_A}}\ket{-\sqrt{\mu}e^{i\delta_b}}_{\mathrm{C_B}}+\ket{-\sqrt{\mu}}_{\mathrm{C_A}}\ket{\sqrt{\mu}e^{i\delta_b}}_{\mathrm{C_B}}\}\bigg]\nonumber\\
	&=e^{-2\mu}\cosh(2\mu).
\end{align}
 Here $P\{\ket{\phi}\}$ represents the density matrix for a quantum state $\ket{\phi}$ (not necessarily normalized). 
 	Recall that the estimation of $e_{\mathrm{ph}}$ is just to count how many pairs of AB in those $K_0$ successful trials where Alice and Bob both chose the code mode are found in either state $\ket{+}_\mathrm{A}\ket{+}_\mathrm{B}$ or $\ket{-}_\mathrm{A}\ket{-}_\mathrm{B}$. Here, we treat the events $\ket{+}_\mathrm{A}\ket{+}_\mathrm{B}$ and $\ket{-}_\mathrm{A}\ket{-}_\mathrm{B}$ as phase errors since $\ket{+}_\mathrm{A}\ket{+}_\mathrm{B}$ and $\ket{-}_\mathrm{A}\ket{-}_\mathrm{B}$ correspond to the twin-field states with even numbers of sum photons and the probability of occurrence of even photon number states is usually lower when attack is absent. In this sense, if Alice and Bob are able to prepare $\rho_{\mathrm{even}}$, $e_{\mathrm{ph}}$ can be estimated by observing its yield. Unfortunately, the preparation of $\rho_{\mathrm{even}}$ is a challenging task.

As the second-best plan, \cite{RN87} offers how we can design the decoy mode to estimate the yield of $\rho_{\mathrm{even}}$. We use a linear combination $\sum_{i}\alpha^{(i)}\rho^{(i)}$ ($\alpha^{(i)}\in\mathbb{R}$) of decoy states $\{\rho^{(i)}\}$ to approximate $\rho_{\mathrm{even}}$, namely
	\begin{equation}
		p_{10}^2\tau(\mu_0)+p_{11}^2\frac{\Delta}{\pi}\tau(\mu_1)-\Gamma\tau(\mu_2)\geqslant\Lambda\rho_{\mathrm{even}}\label{ineq},
	\end{equation}
	which will be proved later on. The inequality (\ref{ineq}) is called operator dominance inequality. Here, $p_{10}^2$ is the probability that Alice and Bob both send the vacuum state in a trial. $p_{11}^2\frac{\Delta}{\pi}$ is the probability that Alice and Bob both send the weak coherent state with intensity $\mu_1$ and the random phase $\theta_a$ and $\theta_b$ satisfies $|\theta_a-\theta_b|\ {\rm mod}\ \pi \le\frac{\Delta}{2}$ in a trial. $\tau(\mu_0)=P\{\ket{0}_{\mathrm{C_A}}\ket{0}_{\mathrm{C_B}}\}$, $\ket{0}$ represents the vacuum state.
	\begin{align}
		\tau(\mu)&=\frac{1}{2}\cdot\frac{1}{2\pi}\int_{0}^{2\pi}d\theta\frac{1}{\Delta}\int_{-\frac{\Delta}{2}}^{\frac{\Delta}{2}}P\{\ket{\sqrt{\mu}e^{i\theta}}_{\mathrm{C_A}}\ket{\sqrt{\mu}e^{i(\theta+\delta_b)}}_{\mathrm{C_B}}\}\mathrm{d}\delta_b\nonumber\\
		&+\frac{1}{2}\cdot\frac{1}{2\pi}\int_{0}^{2\pi}d\theta\frac{1}{\Delta}\int_{-\frac{\Delta}{2}}^{\frac{\Delta}{2}}P\{\ket{\sqrt{\mu}e^{i\theta}}_{\mathrm{C_A}}\ket{-\sqrt{\mu}e^{i(\theta+\delta_b)}}_{\mathrm{C_B}}\}\mathrm{d}\delta_b \ (\mu=\mu_1\ \text{or}\ \mu_2)
	\end{align} 
	 is the joint quantum state that satisfies the phase postselection condition in decoy mode when Alice and Bob both choose the same optical pulse intensity $\mu_1$ or $\mu_2$. We choose suitable values of $\Gamma$ and $\Lambda$ as follows: 
	 \begin{align}
	 	\frac{\Gamma}{p_{11}^2}&=\frac{\mu_1e^{-2\mu_1}\Delta}{\mu_2e^{-2\mu_2}\pi},\label{Gamma}\\
	 	\frac{p_{\mathrm{even}}}{\Lambda}&=\sum_{s\in \mathbb N_{\mathrm{even}}}P^\mu_s\cdot q_s^{-1},\label{peven/Lambda}
	 \end{align}
	 where
	 \begin{align}
	 	P^\mu_s&=e^{-2\mu}\frac{(2\mu)^s}{s!},\\
	 	q_s&=\
	 	\begin{cases}
	 		p_{10}^2-p_{11}^2\frac{\Delta}{\pi}e^{-2\mu_1}(\mu_1-\mu_2)/\mu_2& s=0\\
	 		p_{11}^2\frac{\Delta}{\pi}\mu_1e^{-2\mu_1}2^s(\mu_1^{s-1}-\mu_2^{s-1})/s!& s\geq 1,
	 	\end{cases}\label{qs}
	 \end{align}
 and $\mathbb N_{\mathrm{even}}$ is the set of nonnegative even numbers.
	 
	 Indeed, the inequality (\ref{ineq}) can be rewritten as the equation
	 \begin{equation}
	 	p_{10}^2\tau(\mu_0)+p_{11}^2\frac{\Delta}{\pi}\tau(\mu_1)=\Gamma\tau(\mu_2)+\Lambda\rho_{\mathrm{even}}+T\rho_{\mathrm{junk}}\label{eq},
	 \end{equation}
 where $\rho_{\mathrm{junk}}$ is a normalized useless `junk' state and $T:=p_{10}^2+p_{11}^2\frac{\Delta}{\pi}-\Gamma-\Lambda$.
 
 In the actual protocol, if Alice and Bob both choose the decoy mode and send states with intensity $\mu_0$ or send states with intensity $\mu_1$ which satisfy the phase postselection condition $|\theta_a-\theta_b|\ {\rm mod}\ \pi \le\frac{\Delta}{2}$, we define it as decoy1 mode. If they both choose decoy mode and send states with intensity $\mu_2$ which satisfy the phase postselection condition, it is a decoy2 mode. If they both choose code mode, it is a code0 mode. 
 
 Obviously, $K_0$ is the detection frequency of code0 mode and $K_2$ is the detection frequency of decoy2 mode under this definition. What's more, the total detection frequency $K_0$ can be decomposed into two parts with and without phase error $K_0^{(\mathrm{even})}$ and $K_0-K_0^{(\mathrm{even})}$ in the code0 mode. 
 
\begin{figure}
 	\centering
 	\includegraphics[width=\textwidth]{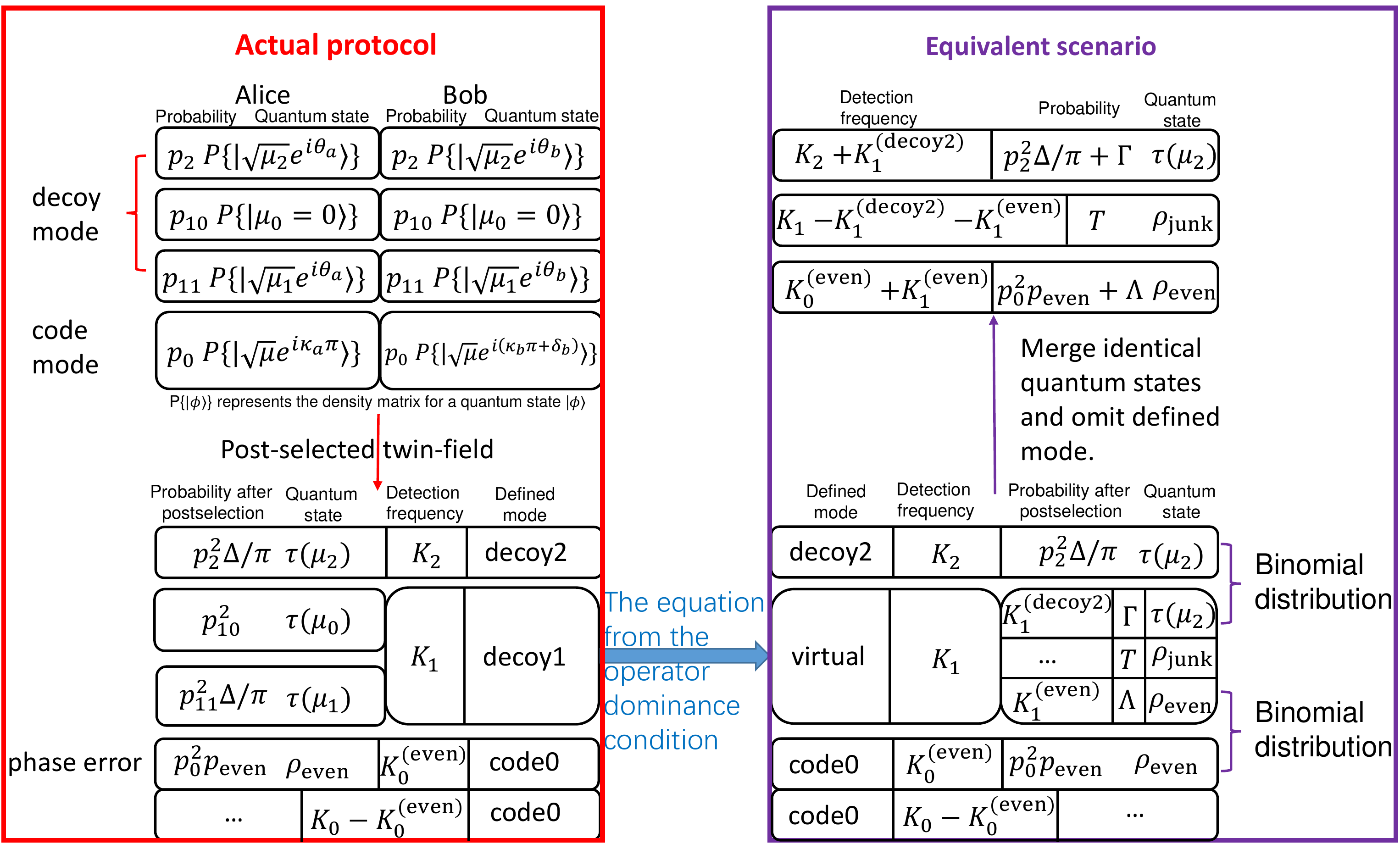}
 	\caption{
 		Relation between the actual protocol and the equivalent scenario. In the actual protocol, Alice or Bob chooses code mode with probability $p_0$ and prepares the weak coherent state $\ket{\sqrt{\mu}e^{i\kappa_a\pi}}$ (or $\ket{\sqrt{\mu}e^{i(\kappa_b\pi+\delta_b)}}$). They prepare vacuum state with $p_{10}$ or a weak coherent state with intensity randomly choosing from $\mu_1$ and $\mu_2$ with probability $p_{11}$ and $p_2$. After postselection, the twin-filed quantum state and the corresponding probability are shown in the figure. The detection frequency is the number of times Charlie has declared success. The defined mode is for the twin-field which is different from the one-party code or decoy mode. If Alice and Bob both choose code mode, the corresponding twin-field is defined as code0 mode, the detection frequency of which is $K_0$. We only focus on $\rho_{\mathrm{even}}$ that causes the phase error. We denote the detection frequancy of $\rho_{\mathrm{even}}$ as $K_0^{\mathrm{(even)}}$. The security of the actual protocol is quantitatively assured if a good upper bound on $K_0^{\mathrm{(even)}}$ is found. To find such a bound, we construct an equivalent scenario of the actual protocol by replacing decoy1 mode with virtual mode which is the mixture of three states $\rho_{\mathrm{even}}$, $\tau(\mu_2)$ and useless `junk' state $\rho_{\mathrm{junk}}$ with probabilities $\Lambda$, $\Gamma$ and $T$ based on the equation from the operator dominance condition. In the equivalent scenario, Alice and Bob prepare twin-field $\rho_{\mathrm{even}}$ which comes from code0 mode with probability $p_0^2p_{\mathrm{even}}$ or from virtual mode with $\Lambda$, prepare $\tau(\mu_2)$ which comes from decoy2 mode with $p_2^2\Delta/\pi$ or from virtual mode with $\Gamma$. Merging identical quantum states and omitting the defined mode, Alice and Bob prepare twin-field $\rho_{\mathrm{even}}$ with probability $p_0^2p_{\mathrm{even}}+\Lambda$, $\tau(\mu_2)$ with probability $p_2^2\frac{\Delta}{\pi}+\Gamma$ and $\rho_{\mathrm{junk}}$ with probability $T$.
 	}
 	\label{fig.equivalent}
 \end{figure}

 Note that in $N_{\text{tot}}$ rounds of actual trials, the detection frequency of the twin-field quantum state $\tau(\mu_0)$ is $K_{10}$ and the detection frequency of $\tau(\mu_1)$ is $K_{11}$, so the detection frequency of decoy1 mode is $K_1=K_{10}+K_{11}$. The equation (\ref{eq}) implies that decoy1 mode can be equivalently decomposed into the mixture of three states $\rho_{\mathrm{even}}$, $\tau(\mu_2)$ and useless `junk' state $\rho_{\mathrm{junk}}$ with probabilities $\Lambda$, $\Gamma$ and $T$. We define the mixture of the three states as virtual mode, so we can construct an equivalent scenario of the actual protocol by replacing decoy1 mode with virtual mode in Fig.\ref{fig.equivalent}.
 
 The equivalent scenario is a combination of code0 and decoy2 mode in the actual protocol and a virtual mode which is equivalent to decoy1 mode of the actual protocol.
 In the equivalent scenario, merging identical quantum states and omitting the defined mode, Alice and Bob send twin-field $\rho_{\mathrm{even}}$ with probability $p_0^2p_{\mathrm{even}}+\Lambda$, $\tau(\mu_2)$ with probability $p_2^2\frac{\Delta}{\pi}+\Gamma$ and $\rho_{\mathrm{junk}}$ with probability $T$. 
 
 We denote the detection frequency of $\tau(\mu_2)$ from the virtual mode as $K_1^{\mathrm{(decoy2)}}$ and $\rho_{\mathrm{even}}$ from the virtual mode as $K_1^{\mathrm{(even)}}$. For twin-field $\tau(\mu_2)$ successfully detected by Charlie, it comes from decoy2 mode with probability $\frac{p_2^2\Delta/\pi}{p_2^2\Delta/\pi+\Gamma}$ and virtual mode with probability $\frac{\Gamma}{p_2^2\Delta/\pi+\Gamma}$. So the detection frequency $K_2$ is a Bernoulli sampling from a population with $K_2+K_1^{\mathrm{(decoy2)}}$ elements, which leads to a lower bound on $K_1^{\mathrm{(decoy2)}}$ in terms of $K_2$. This is also the case with $K_1^{(\mathrm{even})}$ and $K_0^{(\mathrm{even})}$, which leads an upper bound on $K_0^{\mathrm{(even)}}$ in terms of $K_1^{\mathrm{(even)}}$. In an asymptotic limit of $K_1$, $K_2\to\infty$, a bound on $K_0^{(\mathrm{even})}$ is immediately obtained from the relations $K_1^{\mathrm{(decoy2)}}/K_2=\Gamma/(p_2^2\Delta/\pi)$, $K_0^{(\mathrm{even})}/K_1^{(\mathrm{even})}=p_0^2p_{\mathrm{even}}/\Lambda$ and $K_1\geq K_1^{(\mathrm{even})}+K_1^{\mathrm{(decoy2)}}$.
 In the finite size, \cite{RN87} gives the method for constructing $f(K_1,K_2)$ which satisfies
 \begin{equation}
 	\mathrm{Prob}\{K_0^{(\mathrm{even})}\le f(K_1,K_2)\}\geq 1-\epsilon\label{phaseerrorbound}
 \end{equation}
to estimate the upper bound of $K_0^{(\mathrm{even})}$ with a failure probability $\epsilon$. In the original NPP-TF-QKD finite-key analysis, the upper bound on the number of phase errors is constructed in \cite{RN87} by the use of the Chernoff bound. Here we take $f(K_1,K_2)$ in our protocol (with the corresponding modification) as follows:
\begin{equation}
	f(K_1,K_2)=M^{+}(K_1^{(\mathrm{even})+};\frac{\Lambda}{p_0^2p_{\mathrm{even}}+\Lambda},\frac{\epsilon}{2})
\end{equation}
with
\begin{equation}
	K_1^{(\mathrm{even})+}:=K_1-M^{-}(K_2;\frac{p_2^2\Delta/\pi}{p_2^2\Delta/\pi+\Gamma},\frac{\epsilon}{2}).
\end{equation}
Using the approximate formula for the general bound $M^{\pm}(K;p,\epsilon)\approx\frac{1-p}{p}K\pm\sqrt{-\log\epsilon}\frac{\sqrt{2(1-p)}}{p}\sqrt{K}$ when $(1-p)K\gg-\log\epsilon$ in \cite{RN87}, we can approximate $f(K_1,K_2)$ as
\begin{equation}
	f(K_1,K_2)=\frac{p_0^2 p_{\mathrm{even}}}{\Lambda}\biggl(K_1-\frac{\Gamma\pi}{p_2^2\Delta}K_2+\nu(K_1,K_2)\sqrt{-\log(\epsilon/2)}\biggr),
\end{equation}
where $\nu(K_1,K_2)=\frac{\sqrt{2\Gamma(p_2^2\frac{\Delta}{\pi}+\Gamma)}}{p_2^2\frac{\Delta}{\pi}}\sqrt{K_2}+\sqrt{2(1+\frac{\Lambda}{p_0^2p_{\mathrm{even}}})}\sqrt{K_1-\frac{\Gamma}{p_2^2\frac{\Delta}{\pi}}K_2}$. So we can estimate the upper bound of $e_{\mathrm{ph}}$ by $e_{\mathrm{ph}}\leq \frac{f(K_1,K_2)}{K_0}$.


 It is known \cite{RN127, RN87} that if the number of phase errors is bounded with a failure probability $\epsilon$ as in Eq.(\ref{phaseerrorbound}), the protocol  with the final key length $G=K_0-\lceil K_0 h(f(K_1,K_2)/K_0)\rceil-H_{\mathrm{EC}}-\zeta-\zeta^{\prime}$ is $\epsilon_{\mathrm{sct}}$-secret and  $\epsilon_{\mathrm{cor}}$-correct. 
	Here,  $\epsilon_{\mathrm{sct}}=\sqrt{2}\sqrt{\epsilon+2^{-\zeta}}$ \cite{Hayashi_2012} means the final key is secret except a failure probability no more than $\epsilon_{\mathrm{sct}}$, $\zeta$ is the 
	number of additional bits consumed in the universal$_2$-hash-function-based privacy amplification to take the failure probability of privacy amplification itself into account \cite{Sheridan_2010} and $h(x)=-x{\log_2}(x)-(1-x){\rm log_2}(1-x)$ is the binary Shannon entropy function. Meanwhile, $\zeta^{\prime}$ bits are consumed to ensure that the failure probability of error verification is at most to $\epsilon_{\mathrm{cor}}=2^{-\zeta^{\prime}}$, so the protocol is $\epsilon_{\mathrm{cor}}$-correct, i.e. the probability for Alice's and Bob's final key to differ is bounded by $\epsilon_{\mathrm{cor}}$. Finally, our protocol becomes $\epsilon_{\mathrm{sec}}$-secure with $\epsilon_{\mathrm{sec}}=\epsilon_{\mathrm{sct}}+\epsilon_{\mathrm{cor}}$ in the sense of the composable security definition \cite{RN110}. 

	To prove (\ref{ineq}), we divide the operator dominance inequality into an in-phase part and an anti-phase part, say, 
	\begin{align}
		p_{\mathrm{even}}\rho_{\mathrm{even}}^+=\frac{1}{\Delta}\int_{-\frac{\Delta}{2}}^{\frac{\Delta}{2}}\frac{1}{4}P\{\ket{\sqrt{\mu}}_{\mathrm{C_A}}\ket{\sqrt{\mu}e^{i\delta_b}}_{\mathrm{C_B}}+\ket{-\sqrt{\mu}}_{\mathrm{C_A}}\ket{-\sqrt{\mu}e^{i\delta_b}}_{\mathrm{C_B}}\}\mathrm{d}\delta_b\label{rho+},\\
		p_{\mathrm{even}}\rho_{\mathrm{even}}^-=\frac{1}{\Delta}\int_{-\frac{\Delta}{2}}^{\frac{\Delta}{2}}\frac{1}{4}P\{\ket{\sqrt{\mu}}_{\mathrm{C_A}}\ket{-\sqrt{\mu}e^{i\delta_b}}_{\mathrm{C_B}}+\ket{-\sqrt{\mu}}_{\mathrm{C_A}}\ket{\sqrt{\mu}e^{i\delta_b}}_{\mathrm{C_B}}\}\mathrm{d}\delta_b,\\
		\tau^+(\mu)=\frac{1}{2\pi}\int_{0}^{2\pi}\mathrm{d}\theta\frac{1}{\Delta}\int_{-\frac{\Delta}{2}}^{\frac{\Delta}{2}}P\{\ket{\sqrt{\mu}e^{i\theta}}_{\mathrm{C_A}}\ket{\sqrt{\mu}e^{i(\theta+\delta_b)}}_{\mathrm{C_B}}\}\mathrm{d}\delta_b\label{tau+}\ (\mu=\mu_1\ or\ \mu_2),\\
		\tau^-(\mu)=\frac{1}{2\pi}\int_{0}^{2\pi}\mathrm{d}\theta\frac{1}{\Delta}\int_{-\frac{\Delta}{2}}^{\frac{\Delta}{2}}P\{\ket{\sqrt{\mu}e^{i\theta}}_{\mathrm{C_A}}\ket{-\sqrt{\mu}e^{i(\theta+\delta_b)}}_{\mathrm{C_B}}\}\mathrm{d}\delta_b\ (\mu=\mu_1\ or\ \mu_2).
	\end{align}
	So we only need to prove the following operator dominance inequalities:
	\begin{align}
		p_{10}^2\tau(\mu_0)+p_{11}^2\frac{\Delta}{\pi}\tau^+(\mu_1)-\Gamma\tau^+(\mu_2)\geqslant\Lambda\rho_{even}^+\label{ineq+},\\
		p_{10}^2\tau(\mu_0)+p_{11}^2\frac{\Delta}{\pi}\tau^-(\mu_1)-\Gamma\tau^-(\mu_2)\geqslant\Lambda\rho_{even}^-\label{ineq-}.
	\end{align}
	Now we give the method to calculate $\Gamma$ and $\Lambda$ from the parameters $(\mu,\mu_0,\mu_1,
	\mu_2,p_{10},p_{11},\Delta)$. The state of pulse pair when Alice sends the coherent state $\ket{\sqrt{\mu}e^{i\theta_a}}$ and Bob sends the coherent state $\ket{\sqrt{\mu}e^{i\theta_b}}$ is
	\begin{align}
		\ket{\sqrt{\mu}e^{i\theta_a}}\ket{\sqrt{\mu}e^{i\theta_b}}&=e^{-\frac{\mu}{2}}\sum_{m=0}^\infty\frac{(\sqrt{\mu}e^{i\theta_a})^m}{\sqrt{m!}}\ket{m}e^{-\frac{\mu}{2}}\sum_{n=0}^\infty\frac{(\sqrt{\mu}e^{i\theta_b})^n}{\sqrt{n!}}\ket{n}\nonumber\\
		&=e^{-\mu}\sum_{m,n=0}^{\infty}\frac{(\sqrt{\mu})^{m+n}e^{i(m\theta_a+n\theta_b)}}{\sqrt{m!n!}}\ket{m}\ket{n}\nonumber\\
		&=\sum_{s=0}^{\infty}\sqrt{P^\mu_s}\sum_{m=0}^{s}\ket{m,s-m}_{\theta_a,\theta_b}\nonumber\\
		&=\sum_{s=0}^{\infty}\sqrt{P^\mu_s}\ket{s}_{\theta_a,\theta_b}.
	\end{align}
	Here, $\ket{m,s-m}_{\theta_a,\theta_b}=\sqrt{\frac{s!}{2^sm!(s-m)!}}e^{im\theta_a}e^{i(s-m)\theta_b}\ket{m}\ket{s-m}$, $\ket{s}_{\theta_a,\theta_b}=\sum_{m=0}^{s}\ket{m,s-m}_{\theta_a,\theta_b}$ is the normalized $s$-photon twin-field state when Alice's phase is $\theta_a$, Bob's phase is $\theta_b$ and their total photon number is $s$. Similarly,
	\begin{equation}
		\ket{-\sqrt{\mu}e^{i\theta_a}}\ket{-\sqrt{\mu}e^{i\theta_b}}=\sum_{s=0}^{\infty}(-1)^s\sqrt{P^\mu_s}\ket{s}_{\theta_a,\theta_b}.
	\end{equation}

    Additionally, we assume these variables are subject to
    \begin{align}
    	0<\mu_2\leq\mu_1\leq1\label{constraint1},\\
    	0<\mu\leq1\label{constraint2},\\
    	\frac{\mu_1-\mu_2}{\mu_2}<\frac{p_{10}^2}{p_{11}^2\frac{\Delta}{\pi}e^{-2\mu_1}}\label{constraint3}.
    \end{align}
    
	Next, we will prove why Eq.(\ref{ineq+}) holds under the above parameters. With the sum-photon twin-field states, Eq.(\ref{tau+}) can be rewritten as
	
	\begin{align}
		\tau^+(\mu)
		&=\frac{1}{2\pi}\int_{0}^{2\pi}\mathrm{d}\theta\frac{1}{\Delta}\int_{-\frac{\Delta}{2}}^{\frac{\Delta}{2}} P\{\ket{\sqrt{\mu}e^{i\theta}}_{\mathrm{C_A}}\ket{\sqrt{\mu}e^{i(\theta+\delta_b)}}_{\mathrm{C_B}}\}\mathrm{d}\delta_b\nonumber\\
		&=\frac{1}{2\pi}\int_{0}^{2\pi}\mathrm{d}\theta\frac{1}{\Delta}\int_{-\frac{\Delta}{2}}^{\frac{\Delta}{2}} P\{e^{-\mu}\sum_{m,n=0}^{\infty}\frac{(\sqrt{\mu})^{m+n}e^{i[m\theta+n(\theta+\delta_b)]}}{\sqrt{m!n!}}\ket{m}\ket{n} \}\mathrm{d}\delta_b\nonumber\\
		&=\frac{1}{2\pi}\int_{0}^{2\pi}\mathrm{d}\theta\frac{1}{\Delta}\int_{-\frac{\Delta}{2}}^{\frac{\Delta}{2}} e^{-2\mu}\sum_{m,n=0}^{\infty}\sum_{u,v=0}^{\infty}\frac{(\sqrt{\mu})^{m+n}e^{i[m\theta+n(\theta+\delta_b)]}}{\sqrt{m!n!}}\ket{m}\ket{n}\frac{(\sqrt{\mu})^{u+v}e^{-i[u\theta+v(\theta+\delta_b)]}}{\sqrt{u!v!}}\bra{u}\bra{v} \mathrm{d}\delta_b\nonumber\\
		&=\frac{1}{2\pi}\int_{0}^{2\pi}\mathrm{d}\theta\frac{1}{\Delta}\int_{-\frac{\Delta}{2}}^{\frac{\Delta}{2}} e^{-2\mu}\sum_{s=0}^{\infty}\sum_{m=0}^{s}\sum_{t=0}^{\infty}\sum_{u=0}^{t}\frac{(\sqrt{\mu})^{s}e^{i[s\theta+(s-m)\delta_b]}}{\sqrt{m!(s-m)!}}\ket{m}\ket{s-m}\frac{(\sqrt{\mu})^{t}e^{-i[t\theta+(t-u)\delta_b]}}{\sqrt{u!(t-u)!}}\bra{u}\bra{t-u} \mathrm{d}\delta_b\nonumber\\
		&=\frac{1}{\Delta}\int_{-\frac{\Delta}{2}}^{\frac{\Delta}{2}} e^{-2\mu}\sum_{s=0}^{\infty}\sum_{m=0}^{s}\sum_{t=0}^{\infty}\sum_{u=0}^{t}\frac{1}{2\pi}\int_{0}^{2\pi}e^{i(s-t)\theta}\mathrm{d}\theta\frac{(\sqrt{\mu})^{s}e^{i(s-m)\delta_b}}{\sqrt{m!(s-m)!}}\ket{m}\ket{s-m}\frac{(\sqrt{\mu})^{t}e^{-i(t-u)\delta_b}}{\sqrt{u!(t-u)!}}\bra{u}\bra{t-u} \mathrm{d}\delta_b\nonumber\\
		&=\frac{1}{\Delta}\int_{-\frac{\Delta}{2}}^{\frac{\Delta}{2}} e^{-2\mu}\sum_{s=0}^{\infty}\sum_{m=0}^{s}\sum_{u=0}^{s}\frac{(\sqrt{\mu})^{s}e^{i(s-m)\delta_b}}{\sqrt{m!(s-m)!}}\ket{m}\ket{s-m}\frac{(\sqrt{\mu})^{s}e^{-i(s-u)\delta_b}}{\sqrt{u!(s-u)!}}\bra{u}\bra{s-u} \mathrm{d}\delta_b\nonumber\\
		&=\frac{1}{\Delta}\int_{-\frac{\Delta}{2}}^{\frac{\Delta}{2}} \sum_{s=0}^{\infty}e^{-2\mu}\frac{(2\mu)^s}{s!}\sum _{m=0}^s \sqrt{\frac{s!}{2^s m! (s-m)!}}e^{i(s-m)\delta_b}\ket{m}\ket{s-m}\sum _{u=0}^{s} \sqrt{\frac{s!}{2^s u! (s-u)!}}e^{-i (s-u)\delta _b}\bra{u}\bra{s-u}\mathrm{d}\delta_b\nonumber\\
		&=\frac{1}{\Delta}\int_{-\frac{\Delta}{2}}^{\frac{\Delta}{2}}\sum_{s=0}^\infty e^{-2\mu}\frac{(2\mu)^s}{s!}\sum_{m=0}^{s}\ket{m,s-m}_{0,\delta_b}(\sum_{u=0}^{s}\ket{u,s-u}_{0,\delta_b})^\dagger\mathrm{d}\delta_b\nonumber\\
		&=\frac{1}{\Delta}\int_{-\frac{\Delta}{2}}^{\frac{\Delta}{2}}   \sum_{s=0}^\infty P_s^{\mu}\ket{s}_{0,\delta_b}\bra{s}_{0,\delta_b}\mathrm{d}\delta_b.
	\end{align}
	Meanwhile, Eq.(\ref{rho+}) can be rewritten as	
	\begin{align}
		p_{\mathrm{even}}\rho_{\mathrm{even}}^+
		&=
		\frac{1}{\Delta}\int_{-\frac{\Delta}{2}}^{\frac{\Delta}{2}}
		P\{
			\sum_{s\in \mathbb{N}_{\mathrm{even}}} \sqrt{P_s^\mu}\ket{s}_{0,\delta_b}
		\}\mathrm{d}\delta_b.
	\end{align}
	As a result, the left side of Eq.(\ref{ineq+}) can be rewritten as
	\begin{align}
		\frac{1}{\Delta}\int_{-\frac{\Delta}{2}}^{\frac{\Delta}{2}}
		\biggl(p_{10}^2\ket{0}_{0,\delta_b}\bra{0}_{0,\delta_b}+
		&p_{11}^2\frac{\Delta}{\pi}\sum_{s=0}^\infty P_s^{\mu_1}\ket{s}_{0,\delta_b}\bra{s}_{0,\delta_b}
		-\Gamma\sum_{s=0}^\infty P_s^{\mu_2}\ket{s}_{0,\delta_b}\bra{s}_{0,\delta_b}\biggr)
		\mathrm{d}\delta_b\nonumber\\
		&=\frac{1}{\Delta}\int_{-\frac{\Delta}{2}}^{\frac{\Delta}{2}}\sum_{s=0}^\infty q_s\ket{s}_{0,\delta_b}\bra{s}_{0,\delta_b}\mathrm{d}\delta_b\nonumber\\
		&\geq\frac{1}{\Delta}\int_{-\frac{\Delta}{2}}^{\frac{\Delta}{2}}\sum_{s\in \mathbb{N}_\mathrm{even}}q_s\ket{s}_{0,\delta_b}\bra{s}_{0,\delta_b}\mathrm{d}\delta_b,
	\end{align}
where we use the definition of $q_s$ in Eq.(\ref{qs}), the relation Eq.(\ref{Gamma}) and the inequality constrains (\ref{constraint1}), (\ref{constraint2}), and (\ref{constraint3}). The right side of Eq.(\ref{ineq+}) can be rewritten as
	\begin{align}
		\frac{1}{\Delta}\int_{-\frac{\Delta}{2}}^{\frac{\Delta}{2}}
		\frac{\Lambda}{p_{\mathrm{even}}}
		\sum_{s,t\in \mathbb{N}_\mathrm{even}}
		\sqrt{P_s^\mu}\sqrt{P_t^\mu}\ket{s}_{0,\delta_b}\bra{t}_{0,\delta_b}\mathrm{d}\delta_b.
	\end{align}
	
	The phase fluctuation $\delta_b$ introduced by Bob in code mode just corresponds to the misalignment of phase postselection in decoy mode, so that the integrals on both sides of Eq.(\ref{ineq+}) have the same form. Obviously, Eq.(\ref{ineq+}) is true if the following formula holds
	\begin{equation}
		\sum_{s\in \mathbb{N}_\mathrm{even}}q_s\ket{s}_{0,\delta_b}\bra{s}_{0,\delta_b}
		\geq
		\frac{\Lambda}{p_{even}}\sum_{s,t\in \mathbb{N}_\mathrm{even}}\sqrt{P_s^\mu}\sqrt{P_t^\mu}\ket{s}_{0,\delta_b}\bra{t}_{0,\delta_b} \label{nointineq}.
	\end{equation}
	Furthermore, Eq.(\ref{nointineq}) holds if
	\begin{equation}
		\sum_{s\in \mathbb{N}_\mathrm{even}}\ket{s}_{0,\delta_b}\bra{s}_{0,\delta_b}
		\geq
		\frac{\Lambda}{p_{even}}\sum_{s,t\in \mathbb{N}_\mathrm{even}}\sqrt{P_s^\mu}\sqrt{P_t^\mu}
		\sqrt{q_s^{-1}}\sqrt{q_t^{-1}}
		\ket{s}_{0,\delta_b}\bra{t}_{0,\delta_b}. \label{fineq}
	\end{equation}
	
	To show Eq.(\ref{fineq}) holds, we first consider an arbitrary normalized state $\ket{\phi}$ which is in the sum-photon twin-field quantum space $\mathcal{S}:=\{\ket{\phi}:\ket{\phi}=\sum_{n\geq0}C_n\ket{n}_{0,\delta_b},\sum_{n\geq0}\left|C_n\right|^2=1
	\}$ acting on both sides of it. Then the left side of Eq.(\ref{fineq}) leads to
	\begin{align}
		\bra{\phi}L.S\ket{\phi}=\sum_{n\in \mathbb{N}_\mathrm{even}}C_n C_n^*.
	\end{align}
	For the right side of Eq.(\ref{fineq}), we have
	\begin{align}
		\bra{\phi}R.S\ket{\phi}
		&=\frac{\Lambda}{p_{\mathrm{even}}}\sum_{s,t\in \mathbb{N}_\mathrm{even}}\sqrt{P_s^\mu}\sqrt{P_t^\mu}
		\sqrt{q_s^{-1}}\sqrt{q_t^{-1}}C_s^* C_t\nonumber\\
		&\leq\frac{\Lambda}{p_{\mathrm{even}}}\sqrt{\sum_{s,t\in \mathbb{N}_\mathrm{even}} P_s^\mu P_t^\mu q_s^{-1} q_t^{-1}}\sqrt{\sum_{s,t\in \mathbb{N}_\mathrm{even}} C_s C_s^* C_t C_t^*}\nonumber\\
		&=\frac{\Lambda}{p_{\mathrm{even}}}
		\sqrt{\sum_{s\in \mathbb{N}_\mathrm{even}}P_s^\mu q_s^{-1}}
		\sqrt{\sum_{t\in \mathbb{N}_\mathrm{even}}P_t^\mu q_t^{-1}}
		\sqrt{\sum_{s\in \mathbb{N}_\mathrm{even}}C_s C_s^*}
		\sqrt{\sum_{t\in \mathbb{N}_\mathrm{even}}C_t C_t^*}\nonumber\\
		&=\frac{\Lambda}{p_{\mathrm{even}}}
		\sum_{s\in \mathbb{N}_\mathrm{even}}P_s^\mu q_s^{-1}
		\sum_{s\in \mathbb{N}_\mathrm{even}}C_s C_s^*\nonumber\\
		&=\bra{\phi}L.S\ket{\phi}.
	\end{align}
	L.S and R.S denote the left-hand side and the right-hand side of (\ref{fineq}). Here the Cauchy-Schwartz inequality and (\ref{peven/Lambda}) are used. The full Hilbert space is $\mathcal{H}=\mathcal{S}\oplus\mathcal{S}^\perp$, where $\mathcal{S}^\perp$ is the orthogonal complement of $\mathcal{S}$. An arbitrary normalized state $\ket{\psi}$ in $\mathcal{H}$ can be uniquely decomposed as $\ket{\psi}=\alpha\ket{\phi}+\beta\ket{\chi}$ (note that the $\ket{n}_{0,\delta_b}$ form an orthonormal basis), where $\ket{\phi}$ is a normalized state in $\mathcal{S}$, $\ket{\chi}$ is a normalized state in $\mathcal{S}^\perp$ and $\left|\alpha\right|^2+\left|\beta\right|^2=1$. Now, we consider an arbitrary normalized state $\ket{\psi}$ in $\mathcal{H}$ acting on both side of Eq.(\ref{fineq}):
	\begin{align}
		\bra{\psi}L.S\ket{\psi}=(\alpha^*\bra{\phi}+\beta^*\bra{\chi})L.S(\alpha\ket{\phi}+\beta\ket{\chi})=\left|\alpha\right|^2\bra{\phi}L.S\ket{\phi},\\
		\bra{\psi}R.S\ket{\psi}=(\alpha^*\bra{\phi}+\beta^*\bra{\chi})R.S(\alpha\ket{\phi}+\beta\ket{\chi})=\left|\alpha\right|^2\bra{\phi}R.S\ket{\phi},
	\end{align} 
	so $\bra{\psi}L.S\ket{\psi}\geq\bra{\psi}R.S\ket{\psi}$. This ends the proof of Eq.(\ref{ineq+}).
	Similarly, Eq.(\ref{ineq-}) can also be proved. So we get the parameters $\Lambda$ and $\Gamma$ that satisfy Eq.(\ref{ineq}).

\section{Numerical experiments} \label{Results}
In the final key formula $G=K_0-\lceil K_0 h(f(K_1,K_2)/K_0)\rceil-H_{\mathrm{EC}}-\zeta-\zeta^{\prime}$, we take $H_{\mathrm{EC}}=\lceil f_{\mathrm{cor}}K_0h(e_{\mathrm{bit}})\rceil$ as the cost of error correction, where $f_{\mathrm{cor}}$ is the error correction inefficiency \cite{RN158}, $e_{\mathrm{bit}}$ is the bit error rate in the code mode. We set the error correction inefficiency $f_{\mathrm{cor}}=1.1$ and fix $\epsilon_{\mathrm{sec}}=2^{-31}$ by assuming $\epsilon=2^{-66}$, $\zeta=66$ and $\zeta^{\prime}=32$.
We denote the distance between Alice and Bob as $L$ (in km) and its loss is 0.2dB/km. As a reasonable and simple assumption, we assume that the distance between Alice and Charlie and the distance between Bob and Charlie are the same, and we assume that the properties of Charlie's two detectors are the same. The detection efficiency of Charlie's apparatus is $\eta_d=30\%$, so the channel transmittance from Alice (Bob) to Charlie is  $\eta=10^{\frac{-0.2 L}{2\times10} } \eta _d$. The dark count probability (also the count rate of the vacuum state) of each detector is $p_d=10^{-8}$ per pulse and the intrinsic error rate due to imperfect optical interference visibility is $e_d=0.03$. The typical simulation parameters can be found in \cite{PhysRevX.8.031043,PhysRevA.98.062323,RN90,RN87}. Since the optical pulses from Alice and Bob after postselection have a phase deviation $\delta_b$, the phase mismatch error is $e_{\mathrm{mis}}=\frac{1}{\Delta}\int_{-\frac{\Delta }{2}}^{\frac{\Delta }{2}} \sin ^2(\frac{\delta_b}{2}) \, \mathrm{d}\delta_b$. As a result, the total misalignment (the probability of a photon hitting an erroneous detector) is $e_m=e_d+(1-e_d)e_{\mathrm{mis}}$. We use the linear model mentioned in \cite{PhysRevX.8.031043} and the similar counting rate formulas is also used in \cite{RN97}. In the code mode, the asymptotic counting rate for correct key bit is denoted by $Q_{\mathrm{corr}}=(1-p_d) e^{-2 \eta  \mu  e_m} (1-(1-p_d) e^{-2 \eta  \mu  (1-e_m)})$, and $Q_{\mathrm{err}}=(1-p_d) e^{-2 \eta  \mu  (1-e_m)} (1-(1-p_d) e^{-2 \eta  \mu  e_m})$ for the error key bit. Hence, the total asymptotic counting rate $Q_\mu=Q_{\mathrm{corr}}+Q_{\mathrm{err}}$, $K_0=N_{\mathrm{tot}}p_0^2Q_\mu$ and the error rate for the sifted key is $e_{\mathrm{bit}}=\frac{Q_{\mathrm{err}}}{Q_\mu}$. In the decoy mode, the asymptotic counting rate conditioned on Alice and Bob both preparing $\tau(\mu_i)$ is denoted by $Q_{\mu_i\mu_i}=2(1-p_d)e^{-\eta\mu_i}(1-(1-p_d)e^{-\eta\mu_i})$. We assume that the observed detection frequencies are just equal to their corresponding mean values, which means $K_1/N_{tot}=(p_{11}^2\frac{\Delta}{\pi}Q_{\mu_1\mu_1}+p_{10}^2Q_{\mu_0\mu_0})$, $K_2/N_{tot}=\frac{\Delta}{\pi}p_2^2Q_{\mu_2\mu_2}$. For any communication distance $L$, we optimize the group of parameters $(\mu,\mu_1,\mu_2,p_0,p_{10},p_{11},p_{2},\Delta)$ to maximize the secret key rate $G$.
	
\begin{figure}
	\centering
	\includegraphics[width=0.5\textwidth]{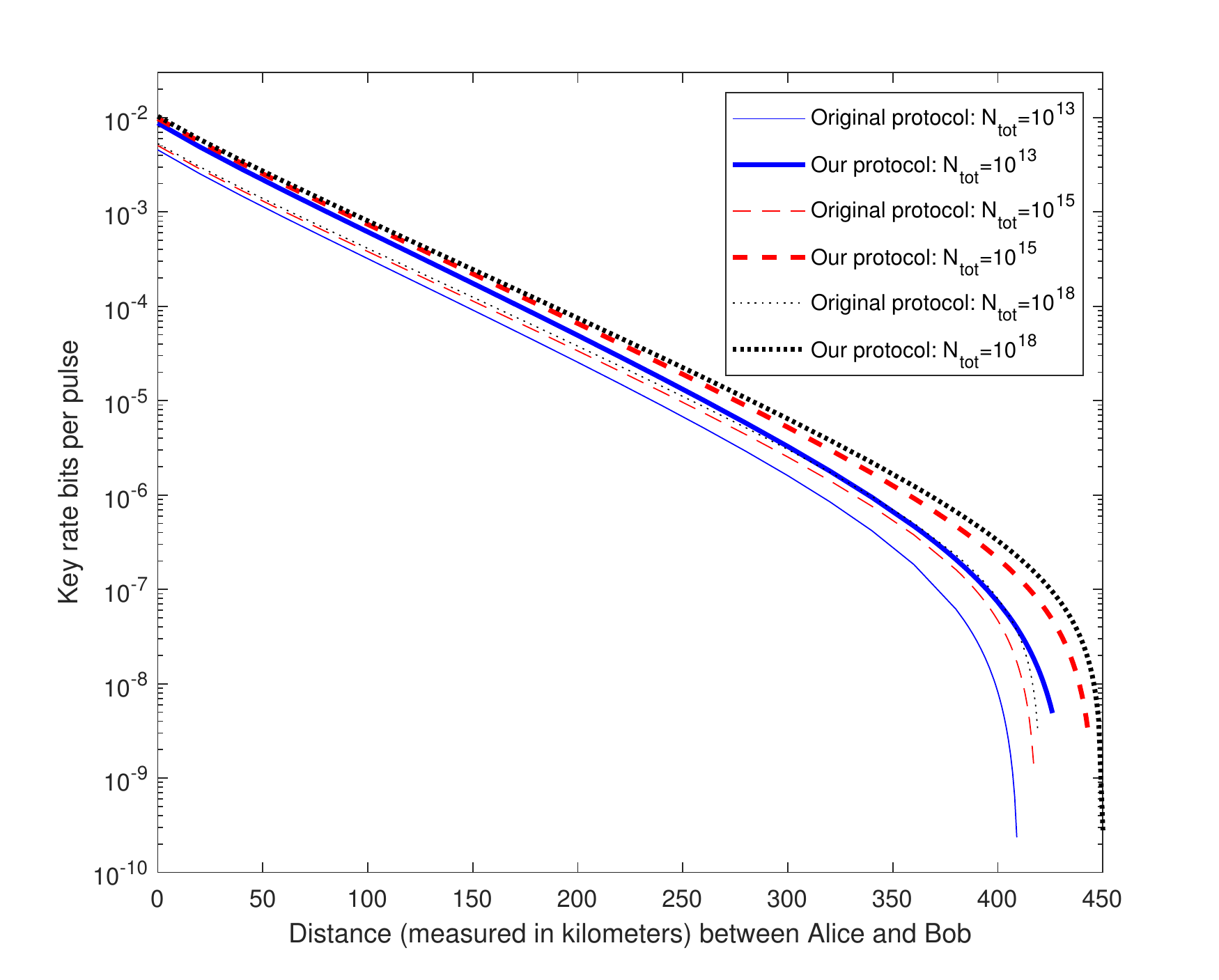}
	\caption{The key rate bits per pulse as a function of distance $L$ (in km) between Alice and Bob. We show the result of our protocol and the original protocol for the $N_{\mathrm{tot}}=10^{13},10^{15}$ and $10^{18}$ cases.} 
	\label{fig.result_1}
\end{figure}

We compare our results with the results \cite{RN87} of the original NPP-TF-QKD protocol in fig.\ref{fig.result_1} for the $N_{\text{tot}}=10^{13},10^{15}$ and $10^{18}$ cases. The achievable distances of our protocol are $426$ km, $443$ km and $450$ km in the three cases, which are $17$ km, $26$ km and $31$ km longer than the original protocol, respectively. At short or medium distances, the key rate bits per pulse of our protocol are nearly doubled compared to original protocol.

\begin{figure}
		\centering
		\includegraphics[width=0.5\textwidth]{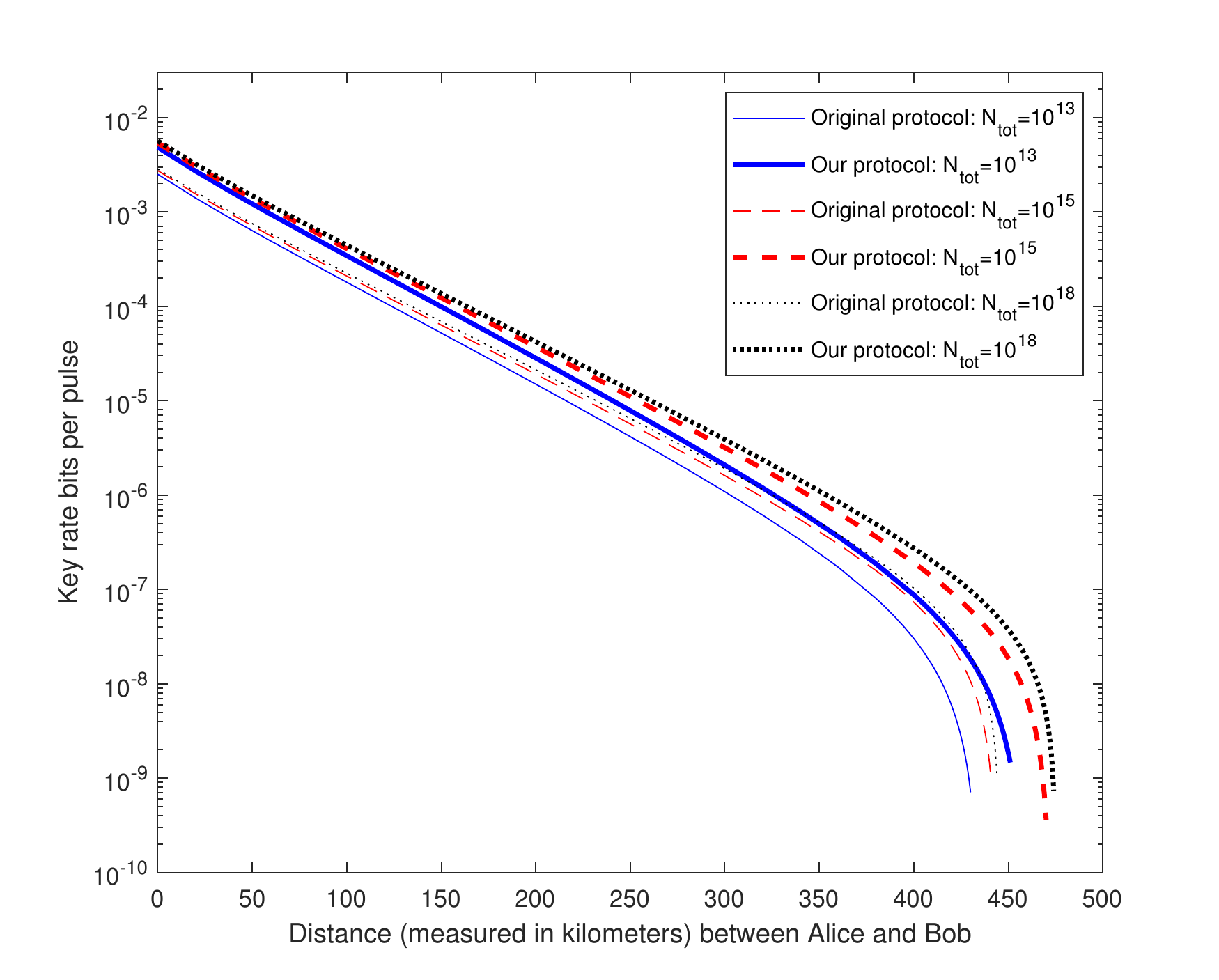}
		\caption{The key rate bits per pulse as a function of distance $L$ (in km) between Alice and Bob in $N_{\mathrm{tot}}=10^{13}$, $10^{15}$ and $10^{18}$ cases for the original protocol \cite{RN97} and our improved protocol. The simulation parameter values are the same as those in this paper. } 
		\label{fig.result_2}
	\end{figure}

Additionally, our idea of partial phase postselection is applicable in other variants of TF-QKD.  Indeed, the four-phase TF-QKD  proposed in \cite{RN97} is very similar to NPP-TF-QKD, except that in code mode both Alice and Bob encode bit in phase from sets $\{0,\pi\}$ or $\{\pi/2,-\pi/2\}$ and then select the rounds with the same coding sets to generate sifted $K_0$ key bits. Evidently, we can introduce the proposed phase postselection to the decoy mode in this protocol. Combined with the security proof in \cite{RN97}, it's easy to prove that $\Gamma$ does not change, but $\Lambda$  is given by 

\begin{equation}
	\frac{p_{\mathrm{even}}}{\Lambda}=\sqrt{\sum_{k,l\in \mathbb{N}_\mathrm{even}\atop k+l=0,4,8...}P^\mu_k P^\mu_l q^{-1}_k q^{-1}_l},
\end{equation}
and
\begin{equation}
	e_{\mathrm{ph}}\le \frac{p_0^2p_{\mathrm{even}}}{2K_0\Lambda}\biggl(K_1-\frac{\Gamma\pi}{p_2^2\Delta}K_2+\nu(K_1,K_2)\sqrt{-\log\frac{\epsilon}{2}}\biggr),
\end{equation}
 where
 \begin{equation}
 	\nu(K_1,K_2)=\frac{\sqrt{2\Gamma\pi(p_2^2\Delta+\Gamma\pi)}}{p_2^2\Delta}\sqrt{K_2}+\sqrt{2(1+\frac{2\Lambda}{p_0^2p_{\mathrm{even}}})}\sqrt{K_1-\frac{\Gamma\pi}{p_2^2\Delta}K_2}.
 \end{equation}
   In the simulation, $K_0/N_{tot}=p_0^2Q_\mu/2$, $K_1/N_{\mathrm{tot}}=(p_{11}^2\frac{\Delta}{\pi}Q_{\mu_1\mu_1}+p_{10}^2Q_{\mu_0\mu_0})$, $K_2/N_{\mathrm{tot}}=\frac{\Delta}{\pi}p_2^2Q_{\mu_2\mu_2}$. 
The improvement results are shown in \ref{fig.result_2}. The achievable distances of the improved four-phase protocol are $451$ km, $470$ km and $474$ km when sending $10^{13}$, $10^{15}$ and $10^{18}$ pulses, which are $21$ km, $29$ km and $30$ km longer than the original four-phase protocol. At short or medium distances, the key rate bits per pulse of the improved protocol are also nearly doubled compared to the original protocol.

In summary, we propose a variation of TF-type QKD with partial phase postselection and give the security proof in the finite-key case. Phase postselection being still free in code mode but introduced in decoy mode significantly improves the key rate and achievable distance. Our idea can also be used in the optimized four-phase twin-field protocol proposed by \cite{RN97}, which confirms its potential advantages in terms of key rate and achievable distance. 
	
\begin{acknowledgments}
This work has been supported by the National Key Research and Development Program of China (Grant No. 2020YFA0309802), the National Natural Science Foundation of China (Grant Nos. 62171424, 61961136004).
\end{acknowledgments}

\bibliography{ref}

\end{document}